\documentstyle[aps,graphicx,epsf]{revtex}\tighten
\def \BE {\begin{equation}}
\def \EE {\end{equation}}
\def \nt {\tau}

\def \ol {\overline}

\def \pw {plane-wave\ }

\def \eq#1 {\eqno{(#1)}}

\def \a {\alpha}

\def \p {\phi}

\def \r {\rho}

\def \e#1 {{\rm e}^{#1}}

\def \ha { { 1\over 2 }}

\def \td {\tilde}

\def \del {\partial}

\begin{document}

\title{Generalizations of pp-wave spacetimes in higher dimensions}
\author{A. Coley\dag, R. Milson\dag, N. Pelavas\dag, V. Pravda\ddag, A. Pravdov\'{a}\ddag ~and R. Zalaletdinov\dag }
\address{\dag\ Department of Mathematics and Statistics,
Dalhousie University, Halifax, Nova Scotia}
\address{\ddag\ Mathematical Institute,
Academy of Sciences, \v Zitn\' a 25, 115 67 Prague 1, Czech Republic}


\maketitle
\begin{abstract}

We shall investigate $D$-dimensional Lorentzian spacetimes
in which all of the~scalar invariants
constructed from the~Riemann tensor and its
covariant derivatives are
zero. These spacetimes are  higher-dimensional
generalizations of $D$-dimensional pp-wave spacetimes,
which have been of interest recently in the context of string theory
in curved backgrounds in higher dimensions.

\end{abstract}

\section{Introduction}

Higher-dimensional pp-wave spacetimes
are of current interest in  string theory
in curved backgrounds, particularly
since these Lorentzian  spacetimes are exact solutions in string theory
and their spectrum can therefore be explicitly determined.
In this paper we shall discuss $D$-dimensional Lorentzian spacetimes
in which all of the~scalar invariants
constructed from the~Riemann tensor and its
covariant derivatives are
zero. These spacetimes can be regarded as  higher-dimensional
generalizations of $D$-dimensional pp-wave spacetimes.

This research follows on from the recent work of \cite{cppm}
in four dimensions, in which it was proven that in
Lorentzian spacetimes
all of the~scalar invariants
constructed from the~Riemann tensor and its
covariant derivatives are
zero if and only if the~spacetime is of Petrov type
III, N or O, all eigenvalues of
the~Ricci tensor are zero and the~common multiple null
eigenvector $l^a$ of the~Weyl and Ricci tensors
is geodesic,  shearfree, non-expanding, and non-twisting
\cite{cppm} (i.e., the  Newman-Penrose (NP) coefficients
$\kappa$, $\sigma$, and $\rho$ are zero); we shall refer to
these spacetimes as vanishing scalar invariant (VSI) spacetimes.
The~Ricci tensor has the~form
\begin{equation}
 R_{a b} = - 2 A l_{a}
 l_{b} + 4 A_i  l_{(a} m^i_{b)} \label{Ricci}
\end{equation}
($i=1,2$).
The~Pleba\' nski-Petrov type (PP-type) is N    for $A_i\not= 0$
or O for $A_i=0$.
We note that for PP-type N, using a null rotation a boost and a spatial
rotation
we can transform away the~Ricci component $A$
and set $A_i=1$. For PP-type O
it is possible to set $A=1$
by performing a boost.

It is known that the energy-momentum tensor for
a spacetime corresponding to  PP-type N
cannot satisfy the weak energy conditions \cite{kramer},
and hence such spacetimes of are not regarded as physical
in classical general relativity (however, see \cite{energy}).  Therefore,  attention is  usually restricted to
PP-type O models, in which the energy-momentum tensor
corresponds to
a pure null radiation field \cite{kramer}.
All of these spacetimes belong to Kundt's class, and hence
the~metric of these spacetimes can be
expressed in an appropriate form  in adapted coordinates \cite{kramer,kundt}.
The  metrics  for
all VSI spacetimes are displayed in  \cite{cppm}.
The generalized pp-wave
solutions are of Petrov-type N, PP-type O
(so that the~Ricci tensor has the~form of null radiation)
with $\nt=0$, and admit a covariantly constant null vector field
\cite{jordan}. The
vacuum spacetimes, which are obtained by setting $A= 0$,
are the~well-known
pp-wave spacetimes (or plane-fronted gravitational
waves with parallel rays).

\section{Higher Order Theorem}

This theorem can be readily generalized to higher dimensions.
We shall study Lorentzian VSI spacetimes in arbitrary $D$-dimensions
(not necessary even,
but $D=10$ is of particular importance from string theory) with signature $D-2$.
In principle we could study other signatures;
for example,  manifolds with signature $D-4$ with $D \ge 5$ may also be of physical interest
        \cite{twotime}.

Let the tetrad  be $l,n,m^1,m^2,\ldots,m^i$ ($l,n$ null with  $l^a l_a= n^a n_a = 0, l^a n_a = 1$, $m^i$ real
and spacelike), so that

\BE g_{a b} = 2l_{(a}n_{b)} +  \delta_{jk} m^j_a m^k_b. \label{tetrad} \EE

Using the notation \BE \{[w_p x_q] [y_r z_s]\} \equiv w_px_q y_r z_s - w_p x_q z_r y_s - x_p w_q y_rz_s + x_p w_q z_r y_s +
y_p z_q w_r x_s - y_p z_q x_r w_s -z_p y_q w_r x_s + z_p y_q x_r w_s, \, \EE if all zeroth order invariants vanish then
there exists a null tetrad (\ref{tetrad}) $l, n, m^i \: (i = 1,\ldots,N=D-2)$ such that \cite{newpaper}
 \BE
R_{abcd}   =   A_i \{[l_a n_b] [ l_c m^i_d]\} + B_{[ij]k}
\{[m^i_a m^j_b]
[l_c m^k_d]\} + C_{(ij)} \{[l_a m^i_b]
[l_cm^j_d]\}.  \label{RRR}
\EE
We still have the freedom to ``choose the frame''
and simplify  further, using boosts,
spins and null rotations,
depending on the algebraic structure of the Ricci and Weyl
tensors [a generalization of Petrov and (Petrov-Plebanski) PP
classifications].

From (\ref{RRR}) we obtain the Ricci tensor: \BE R_{bd} = [-A_i + 2B_{[ij]k}  \delta^{jk}] (l_b m^i_d + l_d
m^i_b) + A l_b l_d, \EE where $A \equiv 2 C_{jk}\delta^{jk}$. We can further simplify $R_{bd}$ depending on its
algebraic type.  If the energy conditions are satisfied \BE A_i - 2B_{[ij]k} \delta^{jk} = 0, \EE we shall refer to
this as type $PP\ol{O}$. In this case we have that \BE R_{bd} = A l_b l_d.  \label{PPO} \EE If this condition is
not satisfied, we can use boosts, spins and null rotations to set $A = 0$, which we shall refer to as type $PP \ol{N}$.

From (\ref{RRR}) we obtain the Weyl tensor: \BE C_{abcd} = \Psi_{i} \{[l_a n_b] [l_c m^i_d] \} +
\Psi_{\{ijk\}} \{[m^i_a m^j_b] [l_c m^k_d]\}  + \Psi_{<i\; j>} \{[ l_a m^i_b] [l_c m^j_d]\},
\EE

where
\begin{eqnarray}
\Psi_{i} = 2\Psi_{\{ijk\}}\delta^{jk} & \equiv & C_{abcd} n^a l^b n^c m^d_{i} = \frac{1}{D-2}[(D-3)A_i+2B_{[ij]k} \delta^{jk}], \nonumber\\
\Psi_{\{ijk\}} & \equiv & \frac{1}{2}C_{abcd} m^a_{i} m^b_{j} n^c m^d_{k} = B_{[ij]k}+\frac{1}{D-2}(A_{[i}\delta_{j]k}-2B_{[im|n|} \delta^{mn}\delta_{j]k}) \nonumber\\
\mbox{and\hspace{0.15in}} \Psi_{<i\; j>} & \equiv & \frac{1}{2}C_{abcd} n^a m^b_{i}n^c m_{j}\!^d = C_{(i\;\! j) } - \frac{1}{2(D-2)} A \delta_{i \;\!j}. \nonumber
\end{eqnarray}
In analogy with the Petrov classification, we shall say that spacetimes with $\Psi_{\{ijk\}} \ne 0$ are of type $P \ol{III}$
(in some instances we can use the remaining tetrad freedom in this case to set $\Psi_{<i\; j>} = 0$).  Spacetimes with
$\Psi_{\{ijk\}} = 0$ will be referred to as of type $P \ol{N}$, in this case we can set $\Psi_{i} = 0$.
Conformally flat spacetimes with $\Psi_{\{ijk\}}=0$ and $\Psi_{<i\; j>}=0$
will be referred to as type $P \ol{O}$. [$\Psi_{\{ijk\}}$ and $\Psi_{<i\; j>}$ are
higher-dimensional analogues of the  complex NP coefficents $\ol{\Psi}_3$ and $\ol{\Psi}_4$ in 4 dimensions.
A comprehensive higher-dimensional Petrov classification, which is not necessary here, will be discussed elsewhere.]

For spacetimes of   type $PP\ol{O}$ and type $P \ol{N}$,
the Ricci tensor is given by (\ref{PPO}) and the Weyl tensor is given by
\BE
C_{abcd} = [C_{(ij)}- \frac{1}{2(D-2)} A \delta_{ij}]
\{[ l_a m^i_b] [l_c m^j_d]\}.\label{PN}
\EE

\section{Generalized Kundt spacetimes}

Using the Bianchi and Ricci identities, it is possible to prove \cite{newpaper} that all curvature invariants
of all orders vanish for spacetimes with Riemann tensor of the form of (\ref{RRR}) that satisfy the following
conditions on the covariant derivative of the uniquely defined  null vector $l_{a;b}$
namely
\[
l^a l_a=0,\quad {l^a}_{;b} l^b=0,\quad {l^a}_{;a}=0,\quad
l_{(a;b)}l^{a;b}=0,\quad l_{[a;b]}l^{a;b}=0.\]
In general, the covariant derivative then has the form
\[
l_{a;b}=
L_{11} l_a l_b
+ L_{1i} l_{a} m^i_{b}
+ L_{i1}  m^i_{a} l_{b}.\]

We are consequently led to study
spacetimes which admit a geodesic, shear-free, divergence-free, irrotational
null congruence $l=\partial_v$, and hence
belong to the ``generalized Kundt'' class in which the metric can be written as
\BE
{\rm d}s^2=-2{\rm d} u [ H {\rm d}u+{\rm d} v
+ W_i {\rm d}x^i] + g_{ij}{\rm d}x^i{\rm d}x^j\ ,\label{dsKundt} \EE
where $i,j,k= 1,\ldots,N$ and the metric functions
\[
H=H(u,v,x^i),\quad W_i =W_i (u,v,x^i),\quad g_{ij}=g_{ij}(u,x^i)\] satisfy the remaining vanishing invariant conditions
and the~Einstein field equations (see \cite{kramer,kundt}). We may, without loss of generality, use the remaining
coordinate freedom (e.g., transformations of the form $x^{i'} = x^{i'}(u,x^j)$) to simplify $g_{ij}$. For the
spacetimes considered here we shall diagonalize $g_{ij}$, and in the particular examples below we shall take
$g_{ij}=\delta_{ij}$. The null tetrad is then \BE l= -\partial_v, ~n=\partial_u-(H+ \frac{1}{2}
W^2)\partial_v+W_i\partial_{x^i}, ~m^i=\partial_{x^i}\ ,\label{Kundttetrada} \EE where $W^2 \equiv W_i W_j
\delta^{ij}$. [Note that in 4D the  uniquely defined null vector given by $l=\partial_v$ is the repeated Weyl
eigenvector.]

All of the exact higher-dimensional solutions will be discussed in detail in \cite{newpaper}.
Let us present a subclass of  type $PP\ol{O}$ and type $P \ol{N}$ exact solutions explicitly here,
in which
the Ricci tensor is given by (\ref{PPO}) and the Weyl tensor is given by (\ref{PN}) in the local coordinates
above, and in which
\BE
g_{ij}=\delta_{ij}, ~~W_1 = - \epsilon \frac{v}{x_1}, ~~ W_i = 0 ~~(i \ne 1),
~~ H = - \epsilon \frac{v^2}{4 {x_1}^2}  + H_0(u,x^k),
 \EE
where $\epsilon=1$ corresponds to the case ``$\tau \ne 0$'' (see \cite{cppm}); higher-dimensional pp-wave
spacetimes have $\epsilon = 0$  (``$\tau = 0$'').
In these spacetimes all of the~scalar invariants
constructed from the~Riemann tensor and its
covariant derivatives are
zero. In the case of vacuum the function $H_0(u,x^k)$ satisfies a differential equation.

A second example of a higher-dimensional VSI spacetime is given by type $P \ol{III}$
(``$\tau = 0$'') solutions
\BE
g_{ij}=\delta_{ij}, ~~W_i = \epsilon W_i(u,x^k),
~~ H = \epsilon v H_1(u,x^k)  + H_0(u,x^k).
 \EE
In general these
spacetimes are of type $PP\ol{N}$ (and the remaining tetrad freedom
can be employed to simplify the metric further). In the
case of type $PP\ol{O}$ (null radiation)
the functions $W_i(u,x^k)$ and $H_1(u,x^k)$ satisfy
additional differential equations.
The higher-dimensional type $P \ol{N}$ pp-wave
spacetimes again occur as a subcase with $\epsilon = 0$
($W_i(u,x^k)=0$, $H_1(u,x^k)$=0).

\section{Discussion}

The VSI~ spacetimes have a number of important physical
applications. In particular, in four dimensions a wide range of VSI~spacetimes (in addition to
the pp-wave spacetimes)
are exact solutions in string theory
to all perturbative orders in the~string tension
(even in of the~presence of the~RR five-form field
strength) \cite{coley} (cf. \cite{amati}).
As a result, these models are expected to provide some hints for
the~study of superstrings on more general backgrounds \cite{GSW}.
String theory in pp-wave backgrounds has been studied by many authors
\cite{tseytlin}, partly in a search for a connection between
quantum gravity and gauge theory dynamics.
Solutions of classical field equations for
which the~counter terms required to regularize
quantum fluctuations
vanish are also of importance because
they offer insights into the~behaviour of the~full quantum theory.
A subclass of Ricci flat VSI 4-metrics,
which includes the~pp-wave spacetimes
and some special Petrov type III or N spacetimes,
have vanishing counter terms up to and including two loops
 and thus   VSI
suffer no quantum corrections
to all loop orders \cite{GG}.

Finding  new   string models
with Lorentzian signature  which are exact in $\a'$ and
whose spectrum  can be explicitly determined
is of great interest in the context of string theory
in curved backgrounds in higher dimensions and, indeed,  higher dimensional
generalizations of  pp-wave  backgrounds
have been considered by a number of authors \cite{tseytlin}.
In particular, it was recently realized
\cite{BMN,russot,BFHP} this
solvability property applies to string models corresponding not
only to the NS-NS but also to certain R-R  \pw
backgrounds. (See also \cite{tpp}, and a  general discussion of pp-waves in
$D=10$ supergravity
appeared in \cite{GuevenAD}).

There is also an interesting connection
between pp-wave backgrounds and gauge field theories.  It is known
that any solution of Einstein gravity admits plane-wave backgrounds in
the~Penrose limit \cite{penrose}.  This was extended to solutions of
supergravities in \cite{G}.  It was shown that the~super-pp-wave
background can be derived by the~Penrose limit from the $AdS_p \times
S^q$ backgrounds in \cite{BFHP}.  The~Penrose limit was recognized to
be important in an exploration of the~AdS/CFT correspondence beyond
massless string modes in \cite{AdS/CFT,BMN}.  Maximally supersymmetric
pp-wave backgrounds of supergravity theories in eleven- and
ten-dimensions have also attracted interest \cite{KG}.

Recently the~idea that our universe is embedded in a higher-dimensional world has
received a great deal of renewed attention \cite{brane}.
Due to the~importance of branes
in understanding the
non-perturbative dynamics of string theories,
a number of classical solutions of branes in the
background of a pp-wave have been studied; in particular
a new brane-world model has been introduced in which the~bulk
solution consists of outgoing plane waves (only) \cite{horo}.

For example, a class of pp-wave string solutions
 with non-constant  NS-NS  or R-R  field strengths,
 which are exact type II
superstring solutions to all orders in $\a'$ since
 all corrections to the leading-order field
equations naturally vanish,
were discussed recently \cite{russot}
 (see also \cite{mam}).
 The metric  ansatz and NS-NS 2-form potential
in 10-dimensional superstring theory is given by
\BE
ds^2=-dudv - K(x^k)du^2+dx_i^2+dy_m^2\ , ~~
B_2= b_m (x^k)\ du\wedge dy_m
\ , ~~ H_3 = \del_i  b_m(x^k)\ dx_i\wedge  du\wedge dy_m\ ,
\EE
where $i=1,...,d$ and $m=d+1,..,8$
(and a dilaton of the form $\p= \r_i x_i + \td \phi(u)$ can be included).
In particular, it was found  \cite{russot} that the only non-zero
component
of the generalized curvature is
\BE
R_{uiuj} = - \ha \del_i \del_j K  - \ha  \del_i b_m \del_j b_m
\ .
\EE
 These solutions are consequently  of  type $PP\ol{O}$ and type $P \ol{N}$
(see (\ref{PPO}) and  (\ref{PN})).
There are several special cases.
For  $b_m =0$  the
standard higher-dimensional generalized  pp-wave solution is recovered
with $K=K_0(x)$ being a harmonic function.
WZW models \cite{nw} result when the $b_m$ are linear,
corresponding to homogeneous \pw backgrounds with constant  $H_3$
field.
The  Laplace equation for $b_m$   can   also be solved
by choosing $b_m$ to be the  real  part
of  complex holomorphic functions.
The R-R counterparts of these string models
are direct analogs of the  pp-wave solution \cite{mam}  supported by
a non-constant 5-form
background. Note  that lifts of  the above solutions to 11
dimensions
belong to a  class of $D=11$ pp-wave backgrounds
 first  considered  in \cite{HullVH}.

{\em Acknowledgements}.
This work was supported, in
  part, by the Natural Sciences and Engineering Research Council of Canada.

\end{document}